\documentclass[fleqn,usenatbib]{mnras}

\usepackage[T1]{fontenc}
\usepackage{ae,aecompl}

\usepackage{graphicx}	
\usepackage{amsmath}	
\usepackage{amssymb}	

\newcommand{\bu}{\mbox{{\boldmath$u$}}} 

\title[Evidence for Large-Scale Solar Convection]
 {Evidence for Large-Scale Subsurface Convection in the Sun}

\author[M. F. Woodard]{
M. F. Woodard,$^{1}$\thanks{E-mail: mfw@nwra.com }
\\
$^{1}$NorthWest Research Associates, Inc., 
3380 Mitchell Lane, Boulder, CO 80301-5410, USA\\
}

\date{Accepted XXX. Received YYY; in original form ZZZ}

\pubyear{2016}

\begin{document}
\label{firstpage}
\pagerange{\pageref{firstpage}--\pageref{lastpage}}
\maketitle

\begin{abstract}
A helioseismic statistical waveform analysis of subsurface flow
was performed on two 720-day time series of SOHO/MDI medium-$l$
spherical-harmonic coefficients. The time series coincide with
epochs of high and low solar activity.
Time-dependent coupling-strength coefficients $b^t_s(n,l)$
of modes of the same radial order $n$ and degree $l$, but different azimuthal order
$m$, were inferred from the waveform analysis. These coefficients are sensitive to
flows and general aspherical structure.
For odd values of $s << l$, the coefficient $b^t_s(n,l)$ measures an average over depth
of the amplitude of one spherical-harmonic $(s,t)$ component of the toroidal flow
velocity field. The depth-dependent weighting function defining the average
velocity is the fractional kinetic energy density in radius of modes of the
$(n,l)$ multiplet.
A mean-square $(n,l)$-dependent flow velocity was inferred from the $b$-coefficients
for $s$ in the range $5$ through $35$ for each $n$ and $l$ in the respective
ranges $1$ through $5$ and $120$ through $149$ for the epochs of high and low activity.
A further averaging, over $l$, yielded a root-mean-square flow velocity
as a function of $n$ for each epoch, which average increases from about
$20\,{\rm m s^{-1}}$ at $n=1$ to $35\,{\rm ms^{-1}}$ at $n=5$. The inferred velocities 
are consistent with (though perhaps do not demand) a cellular pattern of flow 
extending over the vertical range of mode sensitivity, estimated to be about
four percent of the solar radius below the photosphere.
\end{abstract}




\section{Introduction}

Turbulent motions in a sun-like star are believed to play a leading role 
in convecting the star's luminosity through the outer portions of its 
envelope. Turbulent convection is also believed to participate,
along with meridional circulation,
in distributing fluid angular momentum within stellar convection zones,
thereby setting up the differential rotation (or `angular velocity')
within the star
({\it e.g.} \citealp{hansen04,stix04}).
A great deal of effort has been directed toward understanding the interactions
of the Sun's mass flows, which are observed over a great range of spatial scales
\citep{kitch99,durney00,derosa02,kuker05,miesch07}.
The most stringent observational constraints on the internal angular velocity
are provided by whole-Sun oscillation-frequency-splitting measurements
\citep{thompson03,howe09}.
For other large-scale subsurface flows, the methods of local helioseismology
are used \citep{gizon05}.
Stellar structure and evolution calculations have traditionally relied on
the so-called mixing-length theory of convection \citep{bohm58}. In the last few
decades, however, numerical simulations of convection-zone-scale flows have
begun to provide a detailed theoretical picture of the dynamic and
magnetic interiors of stars in general and in particular of the Sun
({\it e.g.} \citealp{miesch05}).

Helioseismic measurements are beginning to reveal deep-seated
large-scale turbulent motion in the Sun.
Unlike the differential rotation and the near-surface meridional flow
\citep{giles97,braun98,gonzalez99,giles00,haber02,hughes03,
zhao04,chou05}, the largest turbulent scales have yet to be well, or even
consistently, characterized.
Large-scale turbulent motions are seen in the first $\sim 15$ Mm below
the photosphere \citep{hathaway00,featherstone06}. Although there does not
appear to be an excess of velocity power at ``giant-cell'' scales, i.e. horizontal
scales comparable to the $\approx 200$ Mm depth of the convection zone,
there is evidence of elongated and persistent large-scale flow structures.
Existing helioseismic observations of convection at greater depths are hard 
to reconcile with one another. 
\cite{hanasoge12}, using time-distance analysis, report only upper limits
on flow speeds at $r/R_\odot=0.96$ and $0.92$, which limits
are at least an order of magnitude smaller than the speeds seen in simulations 
(which more or less agree with mixing-length theory). 
The disparity between theory and observation becomes apparent on horizontal scales
somewhat larger than the $\approx 30\,$ Mm supergranulation scale (corresponding
to angular wavenumber $s \approx 120$)
and increases dramatically with increasing horizontal scale.
The upper limits therefore challenge the
conventional picture of convective energy transport in stars.
However, the time-distance results seem to contradict the more recent ring-diagram
measurements of \cite{greer15}, which are in closer agreement with simulations 
and in fact show the amplitude of turbulent motions increasing with depth 
at depths approaching $r/R_\odot=0.96$.
Supergranulation tracking measurements \citep{hathaway13} suggest large-scale
convective motions as deep as $50$ Mm, with amplitudes similar to those
seen by Greer {\it et al.}

This paper describes a helioseismic analysis of large-scale solar 
subsurface turbulence based on $\approx 4$ years of medium-$l$ Doppler images
from the Michelson Doppler Imager (MDI) on the SOHO spacecraft \citep{scherrer95}.
The analysis of these data takes a global-mode, statistical waveform approach, 
in which time-dependent mode-coupling strengths are first inferred from 
spectral-domain covariance data. The coupling strengths are then used
to estimate the power in subsurface turbulence over the range
$5 \le s \le 35$ of angular wavenumber. 

The data sensitivity model used in the analysis is detailed in 
Section~\ref{sec:forward-model} in the context of a more general sensitivity model. 
The analysis of MDI time series resulting in flow velocity measurements 
is presented in Section~\ref{sec:analysis}, together with analogous measurements 
of near-surface magnetic activity.
Section~\ref{sec:summary} summarizes the findings and discusses their implications
and the prospects for detecting yet deeper flows.

\section{The Forward Model}
\label{sec:forward-model}

\subsection{A general data sensitivity model for helioseismic analysis of flows}
In helioseismic statistical waveform analysis, simple
products of the observed solar oscillation signal
are inverted for the subsurface mass flow velocity and aspherical structure.
In waveform analysis and in other seismic analysis, rigorous forward modeling
is crucial to the accurate retrieval of subsurface conditions.
The data analysis described in the following section used oscillation
signal in the form of coefficients $\varphi^{lm}_{\omega}$ in an expansion
of the observed photospheric Doppler velocity in scalar spherical harmonic
functions, $Y^m_l(\theta,\phi)$, of heliographic colatitude, $\theta$,
and longitude, $\phi$,
and in sinusoidal functions, ${\rm e}^{-{\rm i}\omega t}$, of time, $t$.
The analysis uses {\it covariance} data $\varphi^{l'm'}_{\omega'}\varphi^{lm*}_{\omega}$,
where `$*$' means conjugation, for many combinations of $l,m,\omega,l',m',$ and $\omega'$.
The forward model specifies the dependence of
$E[\varphi^{l'm'}_{\omega'}\varphi^{lm*}_{\omega}]$ on the time-varying interior
flow velocity, ${\bf u}({\bf r},t)$, with ${\bf r}$ denoting spherical polar
coordinates $r$, $\theta$, and $\phi$ and $`E'$ denoting statistical expectation.
In view of the weakness of the flows being investigated, the expectation
$E[\varphi^{l'm'}_{\omega'}\varphi^{lm*}_{\omega}]$ is taken to be the sum
of a zeroth-order term $E[\varphi^{l'm'}_{\omega'}\varphi^{lm*}_{\omega}]_0$,
describing the effect of a spherical reference Sun, and a perturbation
$\delta E[\varphi^{l'm'}_{\omega'}\varphi^{lm*}_{\omega}]$ describing the
effect of large-scale turbulent flows.

To model the effect of deep-seated physical perturbations on waves observed
in the photosphere one needs to consider the behavior of the wave field throughout
a substantial volume of the Sun. For the present analysis the wave field
is represented by frequency-domain mode amplitudes, $a^{\alpha}_{\omega}$,
defined by the expansion
\begin{equation}
 \bxi_{\omega}({\bf r}) =
  \sum_{\alpha} a^{\alpha}_{\omega} \bxi_{\alpha}({\bf r}),
 \label{eq:mode-amp-def}
\end{equation}
where $\bxi_{\omega}({\bf r})$ is the component at frequency $\omega$ and
at position $\bf r$ in the solar interior
of the Lagrangian wave displacement and $\bxi_{\alpha}$ are the displacement
eigenfunctions of the normal modes of the spherical reference
model. The index $\alpha$ refers to the usual ($n,l,m$) indices of global
oscillation modes. (In this paper the convention
$x(t) = \sum_{\omega} x_{\omega}\,{\rm e}^{-{\rm i}\omega\,t}$
is used for the Fourier component $x_{\omega}$ of a time series $x(t)$
of finite duration.)

The present data analysis uses approximate expressions
for the zeroth-order (unperturbed) wave-field covariances
and the flow-induced covariance perturbations that
are analogous to expressions derived for local helioseismic
analysis \citep{woodard06}.
In zeroth order, and for modes that are independently excited,
the expectation $E[a^{\alpha'}_{\omega'}a^{\alpha\,*}_{\omega}]$
is zero unless $\alpha'=\alpha$ and $\omega'=\omega$
and the mode-amplitude spectrum has the lorentzian profile
\begin{equation}
 E[\vert a^{\alpha}_{\omega}\vert^2]_0 =
   A^{\alpha} \vert R^{\alpha}_{\omega}\vert^2.
 \label{eq:wave-cov-0}
\end{equation}
The function
\begin{equation}
  R^{\alpha}_{\omega} \approx -[2\omega_{\alpha}(\omega-\omega_{\alpha}+
   {\rm i}\frac{\gamma_{\alpha}}{2})]^{-1}
 \label{eq:resonant-response}
\end{equation}
represents the complex response of a simple damped oscillator to a unit
harmonic driving force of frequency $\omega$. The parameters $\omega_{\alpha}$
and $\gamma_{\alpha}$ are the resonant frequency and damping rate of mode $\alpha$
and $A^{\alpha}$ is a normalization factor.
For the perturbation, one has
\begin{equation}
 \delta E[a^{\alpha'}_{\omega'}a^{\alpha\,*}_{\omega}]
   = -2\omega (R^{\alpha'}_{\omega'}\,E[\vert a^{\alpha}_{\omega}\vert^2]_0
     + R^{\alpha\,*}_{\omega}\,E[\vert a^{\alpha'}_{\omega'}\vert^2]_0)
     \lambda^{\alpha'}_{\alpha,\omega'-\omega},
 \label{eq:delta-wave-cov}
\end{equation}
where the complex-valued coupling coefficient,
$\lambda^{\alpha'}_{\alpha,\omega}$,
describes the effect of subsurface flows and aspherical structure.
The expression
\begin{equation}
 \lambda^{\alpha'}_{\alpha,\omega} =
 -{\rm i}\int_\odot\mathrm{d} m\;\bxi^*_{\alpha'}\cdot
 ({\bu}_{\omega}\cdot\nabla\bxi_{\alpha})
 \label{eq:coupling-coeff}
\end{equation}
({\it e.g.} \citealp{woodard14}),
in which $\mathrm{d}m$ denotes an element of mass and the integration
is carried out over the entire Sun,
describes the first-order effect of the flow velocity.
Mode-coupling effects of various structural asphericities have
been quantified in \cite{lavely92}. Magnetic activity affects 
stellar oscillations \citep{gizon08} and would be expected
to contribute to the coupling coefficients.
The Hermitian property $\lambda^{\alpha}_{\alpha',-\omega}
 = \lambda^{\alpha'\,*}_{\alpha,\omega}$
was assumed in obtaining the above expression for
$\delta E[a^{\alpha'}_{\omega'}a^{\alpha\,*}_{\omega}]$.
This should be a good approximation, provided that the divergence
of the turbulent mass flux can be neglected in the solar interior
and that the vertical flux can be ignored at the outer turning
points of the observed waves.

Because present-day seismic observations
sample only the Earth-facing hemisphere of the Sun, the mode amplitudes
are not independently observable. More precisely, the response
of the observed signal to the amplitudes is given
approximately by
\begin{equation}
 \varphi^{lm}_{\omega} = \sum_{\alpha}
  L^{lm}_{\alpha}\,a^{\alpha}_{\omega}
  + B^{lm}_{\omega},
 \label{eq:leakage-relation}
\end{equation}
where the leakage matrix $L^{lm}_{\alpha}$ is described, for instance,
in \cite{schou94} and $B^{lm}_{\omega}$ is a background signal.
Expressions for $E[\varphi^{l'm'}_{\omega'}\varphi^{lm*}_{\omega}]_0$
and $\delta E[\varphi^{l'm'}_{\omega'}\varphi^{lm*}_{\omega}]$
follow straightforwardly from the above leakage relation and from
Equations~(\ref{eq:wave-cov-0}) and (\ref{eq:delta-wave-cov}).

As in \citet[hereafter W14]{woodard14}, the mode couplings can be expanded
\begin{equation}
 {\lambda}^{\alpha'}_{\alpha,\sigma} =
 {\lambda}^{n'\ell'm'}_{nlm,\sigma} = \sum_s b^t_{s,\sigma}(nl,n'l')\,
 {\gamma}^{l'sl}_{m'tm},
 \label{eq:coupling-coeff-expansion}
\end{equation}
where, apart from sign factors, the ${\gamma}^{l'sl}_{m'tm}$ are
Clebsch-Gordon coefficients and $t=m'-m$.
The usefulness of this expansion stems from the fact that
$b^t_{s,\sigma}(nl,n'l')$
is sensitive to the vector spherical-harmonic component
of degree $s$ and azimuthal order $t$ of the flow velocity field
at frequency $\sigma$.

\subsection{Modeling flow-dependent couplings of modes within multiplets}
The presence of the resonant factors $R^{\alpha}_{\omega}$ and
$R^{\alpha'}_{\omega'}$ in the expression for wave covariance sensitivity
(Equations~(\ref{eq:delta-wave-cov}), (\ref{eq:wave-cov-0}), and
(\ref{eq:resonant-response})) implies that the
$\varphi^{l'm'}_{\omega'}\varphi^{lm*}_{\omega}$
data can be particularly sensitive to the flow velocity
when $\omega$ and $\omega'$ are close to $\omega_{\alpha=(n,l,m)}$
and $\omega_{\alpha'=(n',l',m')}$, respectively,
for specified $n$ and $n'$.
Therefore, in view of Equation~(\ref{eq:coupling-coeff}),
these covariance data should be especially sensitive
to turbulent frequencies of order
$\sigma = \vert \omega_{\alpha'} - \omega_{\alpha}\vert$.
Giant-cell patterns seen in observations \citep{hathaway15}
and in numerical simulations of solar convection \citep{miesch08} 
have lifetimes of weeks to months. These patterns are thus expected to produce
relatively strong couplings between oscillation modes of frequency
differing by less than $\sim 1\,\mu{\rm Hz}$, in particular between nearly-
degenerate modes of the same ($n,l$) multiplet.

The present data analysis is based on measurements of (near-resonant)
couplings of modes of the same $n$ and $l$.
Replacing the indices $\alpha$, $\alpha'$, and $\omega'$ by ($n,l,m$), ($n',l',m'$),
and $\omega+\sigma$, respectively, in Equation~(\ref{eq:delta-wave-cov}) and using
Equation~(\ref{eq:coupling-coeff-expansion}) one obtains an explicit expression
for the wave-covariance sensitivity to the $b^t_{s,\sigma}(nl,n'l')$ parameters,
for any pair of oscillation modes.
Replacing $m'$ by $m+t$ and $n',l'$ by $n,l$ in the resulting expression
and suppressing the multiplet indices $n,l$,
one obtains the sensitivity relation
\begin{equation}
 \delta E[a^{m+t}_{\omega+\sigma}a^{m\,*}_{\omega}] =
   \sum_s K^{st,\sigma}_{m\omega}\,b^t_{s,\sigma},
 \label{eq:special-delta-wave-cov}
\end{equation}
for modes of one multiplet, where the sensitivity kernel is given by
\begin{equation}
 K^{st,\sigma}_{m\omega} =
  -2\omega (R^{m+t}_{\omega+\sigma}\,E[\vert a^m_{\omega}\vert^2]_0
     + R^{m\,*}_{\omega}\,E[\vert a^{m+t}_{\omega+\sigma}\vert^2]_0)
      \gamma^s_{m+t,t,m}.
 \label{eq:kernel}
\end{equation}

The present analysis uses the approximate expressions given in W14 
for the sensitivity of the $b$-coefficients to the subsurface
velocity appropriate in the limit $s<<l$.
In the low-$s$ approximation, the $b$-coefficients
are sensitive to only the odd-$s$ toroidal components of the flow velocity.
For either odd or even $s$, the toroidal component of spherical indices
($s,t$) has the form 
\begin{equation}
 {\bf T}_{s,\omega}^t = -w_{s,\omega}^t(r)\hat{\bf r}\times
  [\hat{\btheta}\partial_{\theta}+
  \hat{\bphi}\frac{\partial_{\phi}}{\rm{sin}\theta})]Y_s^t(\theta,\phi)
 \label{eq:toroidal-form}
\end{equation}
(Equation 25 of \citealp{lavely92}),
where ($\hat{\bf r}$, $\hat{\btheta}$, $\hat{\bphi}$) are the unit vectors
of the spherical-polar coordinates ($r, \theta, \phi$) and
$Y_s^t$ is the corresponding scalar spherical harmonic function.
As in W14, it is convenient to use the scaled radial function
\begin{equation}
 \tilde{w}^t_{s,\sigma}=\sqrt{\frac{s(s+1)}{2\pi}}
   w^t_{s,\sigma}.
 \label{w-tilde}
\end{equation}
As discussed in that paper, the scaled function
contributes $\vert\tilde{w}^t_{s,\sigma}(r)\vert^2$ to the
mean square flow velocity at radius $r$.

Manipulation of Equations 35 through 47 of W14 gives
\begin{equation}
 b^t_{s,\sigma}(n,l) \approx
  \frac{(-1)^l\,l^{3/2}\,f_s\,\Omega^t_{s,\sigma}(n,l)}{\sqrt{s(s+1)}},
 \label{eq:bstsig-sense}
\end{equation}
for odd $s$, where
$\Omega^t_{s,\sigma}(n,l)$ is an average over $r$ of $\tilde{w}^t_{s,\sigma}(r)/r$,
defined above, and $f_s = (-1)^{(s-1)/2}\,(s!!)^2/s!$.
The $r$-dependent weighting function used to obtain $\Omega^t_{s,\sigma}(n,l)$
is the fractional kinetic energy per unit $r$ of modes of the ($n,l$) multiplet.

\section{Data Analysis and Results}
\label{sec:analysis}

Although the covariance data can be inverted directly
for the flow velocity, the approach taken here was more indirect
in that $b$-coefficients were first extracted from the data, by a
least-squares fitting procedure, then subjected to further analysis.
The $b$-coefficients isolate vector spherical-harmonic and frequency
contributions of the velocity field, thus the analysis was similar
in spirit to a multi-channel deconvolution
(MCD, \citealp{jensen98}).

The signal components $\varphi^{lm}_{\omega}$
were computed for two $720$-day time series of MDI medium-$l$
spherical-harmonic coefficients, for $m=-l,l$ in the $l$ range $120-149$.
The midpoints of the time series occur in mid 1997 and late 2000, 
during epochs of high and low solar magnetic activity.
By Equation~(\ref{eq:leakage-relation}) and
due to the narrowness of the global oscillation frequency
profiles, $\varphi^{lm}_{\omega}$ can be used as a proxy for the mode amplitude
$a^{\alpha}_{\omega} = a^{nlm}_{\omega}$ for ${\omega}$ close to the mode frequency
$\omega_{\alpha}$. More precisely,
$\varphi^{lm}_{\omega} \approx L^{lm}_{nlm}\,a^{nlm}_{\omega}$
and the sensitivity of
$\varphi^{l,m+t}_{\omega+\sigma}\,\varphi^{lm*}_{\omega}$
to the $b$-coefficients is obtained by substituting `$\varphi$' for `$a$'
and `$G$' for `$K$', where $G^{st,\sigma}_{m\omega} =
L^{m+t}_{m+t}\,L^{m}_{m}\,K^{st,\sigma}_{m\omega}$,
in Equation~(\ref{eq:special-delta-wave-cov}).

The form of the sensitivity kernel $G$ depends on parameters,
such as the frequency and line width,
defining the frequency profiles of global oscillation modes.
The mode parameters used for this study were obtained from
a recent re-analysis of MDI spherical-harmonic power spectra
\citep{larson15}. The $m$-dependence of the mode frequencies reflects
the Sun's latitude- and radius-dependent angular velocity referred
to the Sun-orbiting frame of the SOHO spacecraft.
The frequencies, $\sigma$, of turbulent components also apply
to this frame.

The perturbation $\delta E[\varphi^{l,m+t}_{\omega+\sigma}\varphi^{lm*}_{\omega}]$
is the expectation of `residual' covariance data
$\varphi^{l,m+t}_{\omega+\sigma}\varphi^{lm*}_{\omega}
- E[\varphi^{l,m+t}_{\omega+\sigma}\varphi^{lm*}_{\omega}]_0$
and the $b$-parameters were obtained from
a simple linear least-squares fit of the forward model to these data.
The zeroth-order expectation of $\varphi^{l,m+t}_{\omega+\sigma}\varphi^{lm*}_{\omega}$
is obtained from Equations~(\ref{eq:wave-cov-0}) and (\ref{eq:leakage-relation}).
In practice, due to the finite line width of the modes and the fact that
the mode frequencies depend only weakly on $m$, a given mode signal leaks into 
more than one component $\varphi^{lm}_{\omega}$ of the observed signal.
A similar leakage, of the photospheric granulation signal and perhaps other signals 
of solar origin, is expected to occur (via the background term $B^{lm}_{\omega}$ 
of Equation~(\ref{eq:leakage-relation})).
The observed background power is important at low frequency.
As a consequence, $E[\varphi^{l,m+t}_{\omega+\sigma}\varphi^{lm*}_{\omega}]_0$
differs substantially from zero when $t$ is a small integer.
Therefore, in the interest of simplicity, the fitting procedure used only covariances 
for which $\vert t \vert > 4$. 
This restriction precludes the detection of flow components of degree
$s$ less than $5$.

The $b$-parameters for a given $n$ and $l$ were obtained one at a time from 
the covariance data according to
\begin{equation}
 b^t_{s,\sigma} = \frac{\sum_{m\omega} G^{st,\sigma\,*}_{m\omega}\,
  \varphi^{m+t}_{\omega+\sigma}\varphi^{m\,*}_{\omega}}
  {\sum_{m\omega} G^{st,\sigma\,*}_{m\omega}\,G^{st,\sigma}_{m\omega}\,}
 \label{eq:bstsig-fit}
\end{equation}
for $m$ ranging between $-l$ and $l$, subject to the condition
$\vert m+t \vert \le l$, and for $\omega$ lying within one full width of the centroids, 
$\omega_{nlm}$, of the (lorentzian) mode profiles. 
$G^{st,\sigma}_{m\omega}$ is the sensitivity kernel described earlier
in this Section.
The above least-squares estimator is analogous to the one given by
Equation 80 of \cite{woodard06}.
Parameter estimates were obtained for $t = 5$ through $32$
and for $s=t-3$ through $t$ at each $t$. These ranges are optimized for sensitivity
to near-sectoral-harmonic components of giant-cell-scale flows.
The $b^t_{s,\sigma}(n,l)$ were estimated for $\sigma/2\pi$ within about
$2\,\mu{\rm Hz}$ of the expected (rotational-advection) frequency
of flow-velocity features of azimuthal order $t$.
The sampled frequency range was intended to be sufficiently wide to
capture the time-scales of rotationally-advected giant-cell-scale
flow patterns.
For both the low- and high-activity epochs the analysis was performed
on modes of $n = 1$ through $5$ and $l=120$ through $149$.
These modes are collectively sensitive to the subsurface flow velocity
to a depth of about $4 \%$ of the solar radius below the photosphere
({\it e.g.}, Figure 5 of \citealp{jcd02}).

The sensitivity relation (\ref{eq:bstsig-sense}) provides a basis for inverting
the odd-$s$ $b$-coefficients for the toroidal velocity profiles
$w^t_{s,\sigma}(r)$.
But as it was not clear at the outset whether large-scale convective
motions would even be detectable, the next step
in the analysis was simply to establish whether or not a flow
signature is actually present in the measured $b$-coefficients. To this end,
unweighted averages, over $l$ at fixed $n$, of the estimated signed coefficients
$(-1)^l\,b^t_{s,\sigma}(n,l)$
were computed, on the assumption that their expected values vary slowly with $l$.
Given that the sensitivity functions of the signed $b$-coefficients are,
by Equation~(\ref{eq:bstsig-sense}),
slowly varying functions of $l$, this assumption is reasonable
provided that the profiles $w^t_{s,\sigma}(r)$ do not vary
too rapidly with $r$ in the layers explored by the modes analyzed.
Denoting the $l$-averaged signed $b$-coefficients by $\xi^t_{s,\sigma}(n)$,
and in view of the relatively narrow range of $l$ used in the analysis,
one obtains, from Equation~(\ref{eq:bstsig-sense}), the expression
\begin{equation}
 \xi^t_{s,\sigma}(n) \approx l_0^{3/2}\,f_s\,\Omega^t_{s,\sigma}(n,l_0)/
  \sqrt{(s(s+1))},
 \label{eq:xi-sensitivity}
\end{equation}
where $l_0$ is a typical value of $l$ in the measured range.

While the averaging procedure does suppress measurement noise,
the scatter of the individual samples suggests that the signal-to-noise
ratio of the individual $\xi^t_{s,\sigma}(n)$ is rather low.
Since there is no obvious reason to expect correlation in the amplitudes and
phases of different flow components $\Omega^t_{s,\sigma}(n,l_0)$,
further averaging of the signed $b$-coefficients seemed unlikely to improve the
statistics of the flow velocity measurement. Instead, estimates of the power
(squared velocity) of the flow were made, as in the analysis of \cite{hanasoge12}.
The approach exploits the fact that, by Equation~(\ref{eq:xi-sensitivity}),
$\Omega^t_{s,\sigma}(n,l_0)$ makes a positive, though small, contribution to each
$\vert \xi^t_{s,\sigma}(n)\vert^2$  and therefore the statistics of the
$\vert \Omega^t_{s,\sigma}(n,l_0)\vert^2$ measurements can be improved by summing
or averaging the $\vert \xi^t_{s,\sigma}(n)\vert^2$. The noise of
the $\xi^t_{s,\sigma}(n)$ measurements themselves is the main contributor
to $\vert \xi^t_{s,\sigma}(n)\vert^2$ and must be known to determine the
flow power contribution.
(Wave realization noise ({\it e.g.}, \citealp{gizon04}) 
is thought to be the main source of noise in helioseismic measurements.)
To estimate the noise power, the $b$-coefficients themselves were averaged over
$l$ in the same way as the signed coefficients, yielding $\eta^t_{s,\sigma}(n)$
analogous to $\xi^t_{s,\sigma}(n)$.
But because the sensitivity of $b^t_{s,\sigma}(n,l)$ to $\Omega^t_{s,\sigma}(n,l)$
alternates rapidly with $l$, according to Equation~(\ref{eq:bstsig-sense}),
the $\eta^t_{s,\sigma}(n)$ are
expected to have far less sensitivity to the flow velocity than
the $\xi^t_{s,\sigma}(n)$. Accordingly, $\vert \eta^t_{s,\sigma}(n)\vert^2$
was taken to represent the noise contribution
to $\vert \xi^t_{s,\sigma}(n)\vert^2$.

\begin{figure}
   \centerline{\includegraphics[width=\columnwidth]{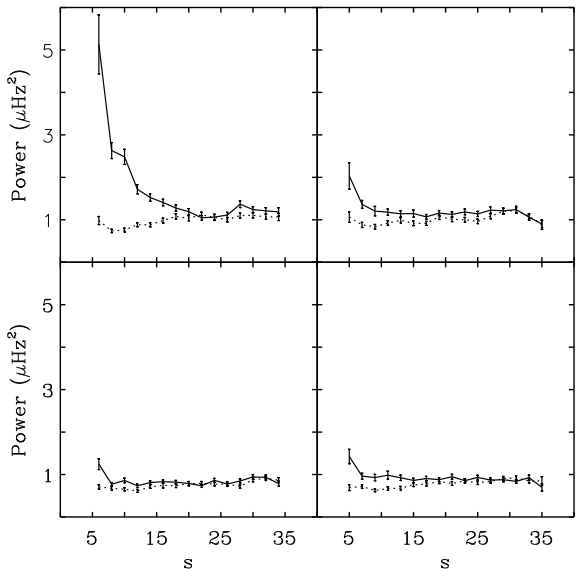}
              }
              \caption{Measures, $\xi^2_s$, of the total power in mode couplings
(solid curves) and of the noise power alone, $\eta^2_s$ (dotted curves),
for even and odd $s$ (left and right panels) and
for high and low solar activity periods (top and bottom panels). As mentioned
in the text, the even-$s$ $\xi^2_s$ are sensitive to magnetic activity while
the odd-$s$ $\xi^2_s$ are sensitive to the subsurface flow velocity.
}
   \label{fig1}
   \end{figure}

For the high- and low-activity epochs the $\xi^t_{s,\sigma}(n)$ and
$\eta^t_{s,\sigma}(n)$ were averaged over $n=1,5$ and the squared moduli
of the resulting averages were then summed over $\sigma$ and averaged over $t$
for the observed values of $s$, $t$, and $\sigma$, yielding $\xi^2_s$ and
$\eta^2_s$. Figure~\ref{fig1} shows $\xi^2_s$ and $\eta^2_s$ as a function of $s$,
for odd and even $s$ separately and for the two epochs.
The errors in quantities derived from the measured $b$-coefficients
are standard deviations based on scatter, where it was assumed that the
$b^t_{s,\sigma}(n,l)$ are statistically independent.
For the even-$s$ case, the measured power, $\xi^2_s$,
displays a striking excess over $\eta^2_s$, the estimated noise power,
near solar maximum. The excess is much smaller near solar minimum,
consistent with the even-s $b$-coefficients being sensitive
to magnetic activity. The analogous excesses for the odd-$s$ case,
though not as striking as the even-$s$ excesses, are statistically significant.
Moreover, the approximate equality of the odd-$s$ excess power
for the high- and low-activity periods does suggest a turbulent,
rather than magnetic, origin for the excess power.

\begin{figure}
   \centerline{\includegraphics[width=\columnwidth]{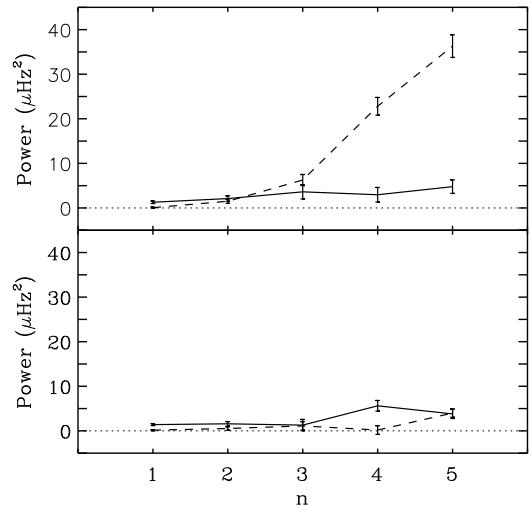}
              }
              \caption{Excess power $\Delta\xi^2(n) = \xi^2(n)-\eta^2(n)$
in mode couplings as a function of $n$ for high- and low-activity periods
(top and bottom panels). Solid curves are a measure of flow velocity power
and dashed curves measure magnetic activity. Errors shown are for the
average curve.
}
   \label{fig2}
   \end{figure}

The $n$-dependence of the power excesses is of great interest since it
should reflect the depth dependence of flow power and solar activity.
Summing $\vert\xi^t_{s,\sigma}(n)\vert^2$ and $\vert\eta^t_{s,\sigma}(n)\vert^2$
over $\sigma$ and averaging the results over observed $t$
yields $\xi^2_s(n)$ and $\eta^2_s(n)$ and the $n$-dependent power excesses
$\Delta\xi^2_s(n) = \xi^2_s(n)-\eta^2_s(n)$. The $n$-dependent excesses
are not as statistically significant as those produced by combining modes of different
$n$. To enhance the statistics of the measurement, the excesses were summed over $s$,
at each $n$, yielding $\Delta \xi^2(n)$. Inspection of Figure~\ref{fig1} suggests 
that the signal-to-noise ratio
of the excess power diminishes for $s$ greater than about $20$, so $s$ values
exceeding $20$ were excluded from the sums.
Figure~\ref{fig2} shows the $n$-dependence of the summed even- and odd-$s$ excesses
for the two epochs. The rapid increase in the even-$s$ excess power with increasing $n$
is consistent with solar activity close to the photosphere \citep{libbrecht90}. 
The much gentler trend in the odd-$s$ excess seems to be consistent with large-scale
turbulent motions over a substantial range of depths.

The depth-averaged flow velocity $\Omega^t_{s,\sigma}(n,l_0)$ can be
expressed in terms of $\xi^t_{s,\sigma}(n)$ using Equation~(\ref{eq:xi-sensitivity}).
Summing the resulting expression for $\vert\Omega^t_{s,\sigma}(n,l_0)\vert^2$
over $t=-s,s$ and $\sigma$ gives
\begin{equation}
 \Omega^2_s(n) \approx s(s+1)(2s+1)\,\Delta\xi^2_s(n)/(l^3_0\,f^2_s)
 \label{eq:ms-velocity}
\end{equation}
for the mean-square depth-averaged flow velocity at degree $s$.
The factor $2s+1$ appears in the above expression because the
quantity $\vert \xi^t_{s,\sigma}\vert^2$, whose $t$-dependence was ignored
for simplicity, had to be summed, rather than averaged, over $t$.
The factor $\Delta\xi^2_s(n)$, rather than $\xi^2_s(n)$, appears in the
preceeding equation to ensure that the right hand expression provides
an unbiased measurement of $\Omega^2_s(n)$.

\begin{figure}
   \centerline{\includegraphics[width=\columnwidth]{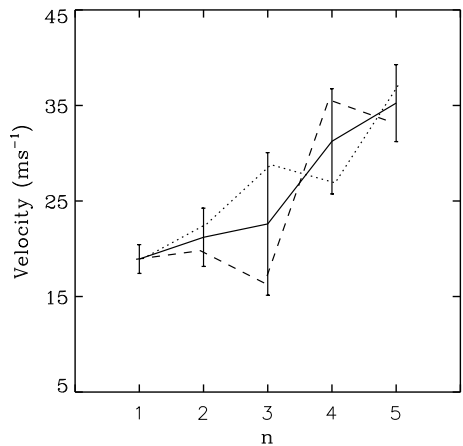}
              }
              \caption{Estimated root-mean-square depth-averaged toroidal flow
velocity, for high- and low-activity periods (dotted and dashed curves)
and their average (solid curve).
}
   \label{fig3}
   \end{figure}

The measured $\Omega^2_s(n)$ were averaged over $s$ at each $n$, with a weighting
inversely proportional to the estimated measurement variance. The averages
were multiplied by $s_{\rm max}=35$ to provide estimates, $\Omega^2(n)$, of the power
in the depth-averaged velocity up to angular wavenumber $s=35$. (It was assumed that
$\Omega^2_s(n)$ can be smoothly interpolated to the unobserved even values
of $s$ and values less than $5$.)
Figure~\ref{fig3} shows the measured root-mean-square depth-averaged linear velocity
$R_\odot\,\Omega(n)$ for the two epochs and the average velocity of the different epochs.

\section{Summary and Outlook}
\label{sec:summary}
The $\sim 30\,{\rm m s^{-1}}$ flow velocities seen in Figure~\ref{fig3}
are roughly consistent with large-scale velocity amplitudes observed
in the photosphere and with the seismic results of \cite{greer15}.
However, they appear to be at odds with the claims of \cite{hanasoge12},
although the present 
measurements do not go as deep as the time-distance measurements. 
(The lower turning radius of the most deeply penetrating, $n=5$, modes 
used for the present analysis, is estimated to be about $0.96 R_\odot$,
corresponding to a depth of about $30$ Mm.)
The modes sample the flow velocity from the photosphere down to their
inner turning radii, with a weighting function that peaks just below
the photosphere. Therefore the modest but apparently significant increase 
in r.m.s. velocity with increasing radial order shown in Figure~\ref{fig3} 
suggests an even greater increase in velocity with physical depth. 
Although an inversion for the depth-dependence of the flow velocity was 
not carried out as part of this analysis,
it seems possible that the present results imply a depth dependence consistent 
with the $100-150\,{\rm m s^{-1}}$ flow velocities at depths approaching $30$ Mm, 
as shown in Figure 4 of Greer {\it et al.}. Note that this
figure is not equivalent to Figure~\ref{fig3} of the current paper, 
because the present analysis is limited to angular wavenumbers
$s \le 35$ while the former includes supergranulation scales.
The supergranulation-tracking analysis of \cite{hathaway13} also suggests
an increase in turbulent power with depth.

In extrapolating the measured flow power, based on near-sectoral 
(i.e., $t \approx s$) flow components, to all $t=-s,s$, strong anisotropy-inducing
effects of solar rotation (seen in numerical convection simulations)
were ignored. Also, the angular wavenumber ($s$) dependence of velocity
power was ignored in computing the overall power. 
In addition, only the toroidal flow velocity was measured.
For these reasons, the overall level of measured turbulent power 
could easily be in error by a factor of $2$ or more. However, unless
the angular anisotropy of large-scale flow patterns changes rapidly
between $r = 0.96 R_\odot$ and the photosphere, the increase in
the r.m.s. velocity with mode depth would seem to indicate a real 
increase in the amplitude of large-scale turbulent motion with increasing depth 
over the sampled depth range, as Greer {\it et al.} found.

That the signal-to-noise ratio of the measured flows is significant even
at the lowest wavenumbers observed suggests the possibility of detecting
turbulent power at angular wavenumbers less than $5$, corresponding to 
the largest angular scales, and at greater depths than the present
analysis permits.  
To measure larger angular scales a better treatment of spatial leakage 
is desirable. To probe deeper into the Sun the analysis would need 
to be extended to oscillation modes of lower $l$.
The signal-to-noise of the measurement can be improved, as only
a fraction of the existing and projected database of solar oscillations 
has been analyzed for this work. Similarly, only the
near-sectoral-harmonic components of the flow have been utilized
in the analysis. The present analysis was based on the dynamical
couplings of modes of the same $n$ and $l$, which are sensitive
mainly to toroidal flow. To detect large-scale poloidal flows
the couplings of modes of different $n$ and/or $l$ would have
to be measured. 

\section*{Acknowledgements}
I thank Curt Cutler and David Hathaway for useful discussions 
and Tim Larson for help in accessing MDI spherical harmonic time series.
I also thank an anonymous referee for constructive comments.
This research was supported by NASA grant
NNX14AH84G to NWRA/CoRA.
This article has been accepted for publication in Monthly Notices of the
Royal Astronomical Society,
Published by Oxford University Press on behalf of the Royal Astronomical Society





\bsp	
\label{lastpage}
\end{document}